\documentclass[10pt,aps,eqsecnum,amsfonts,showkeys]{revtex4}

\usepackage{theorem}
\theorembodyfont{\upshape}

\input epsf

\usepackage{graphicx}

\usepackage{bm}

\newcommand{\beq}{\begin{equation}}
\newcommand{\eeq}{\end{equation}}
\newcommand{\beqs}{\begin{eqnarray}}
\newcommand{\eeqs}{\end{eqnarray}}
\newtheorem{theo}{Theorem}[section]
\newtheorem{cor}{Corollary}[section]
\newtheorem{lemma}[theo]{Lemma}
\newtheorem{defi}{Definition}[section]
\newtheorem{conj}{Conjecture}[section]

\begin{document}

\title{Spanning trees on the Sierpinski gasket}

\author{Shu-Chiuan Chang$^{a,b}$} 
\email{scchang@mail.ncku.edu.tw} 

\author{Lung-Chi Chen$^{c}$} 
\email{lcchen@math.fju.edu.tw}

\bigskip

\affiliation{(a) \ Department of Physics \\
National Cheng Kung University \\
Tainan 70101, Taiwan} 

\bigskip

\affiliation{(b) \ Physics Division \\
National Center for Theoretical Science \\
National Taiwan University \\
Taipei 10617, Taiwan} 

\bigskip

\affiliation{(c) \ Department of Mathematics \\
Fu Jen Catholic University \\
Taipei 24205, Taiwan }

\begin{abstract}

We obtain the numbers of spanning trees on the Sierpinski gasket $SG_d(n)$ with dimension $d$ equal to two, three and four. The general expression for the number of spanning trees on $SG_d(n)$ with arbitrary $d$ is conjectured. The numbers of spanning trees on the generalized Sierpinski gasket $SG_{d,b}(n)$ with $d=2$ and $b=3,4$ are also obtained.


\end{abstract}

\maketitle

%\newpage
%\pagestyle{plain}
%\pagenumbering{arabic}

\section{Introduction}
\label{sectionI}

The enumeration of the number of spanning trees $N_{ST}(G)$ on a graph $G$ was first considered by Kirchhoff in the analysis of electric circuits \cite{kirchhoff}. It is
a problem of fundamental interest in mathematics \cite{bbook,welsh,burton93,lyons05} and physics \cite{temperley,wu77}. The number of spanning trees is closely related to the partition function of the $q$-state Potts model in statistical mechanics \cite{fk,wurev}. Some recent studies on the enumeration of spanning trees and the calculation of their asymptotic growth constants on regular lattices were carried out in Refs. \cite{tzengwu,sw,std,sti}. It is of interest to consider spanning trees on self-similar fractal lattices which have scaling invariance rather than translational invariance. Fractals are geometric structures of noninteger Hausdorff dimension realized by repeated construction of an elementary shape on progressively smaller length scales \cite{mandelbrot,Falconer}. A well-known example of fractal is the Sierpinski gasket which has been extensively studied in several contexts \cite{Gefen80,Gefen81,Rammal,Alexander,Domany,Gefen8384,Guyer,Kusuoka,Dhar97,Daerden,Dhar05}. We shall derive rigorously the numbers of spanning trees on the Sierpinski gasket with dimension equal to two, three and four. The corresponding asymptotic growth constants have simple expressions. We shall also conjecture the general expression of the number of spanning trees on the Sierpinski gasket with arbitrary dimension.

\section{Preliminaries}
\label{sectionII}

We first recall some relevant definitions for spanning trees and the Sierpinski gasket in this section. A connected graph (without loops) $G=(V,E)$ is defined by its vertex (site) and edge (bond) sets $V$ and $E$ \cite{bbook,fh}.  Let $v(G)=|V|$ be the number of vertices and $e(G)=|E|$ the number of edges in $G$.  A spanning subgraph $G^\prime$ is a subgraph of $G$ with the same vertex set $V$ and an edge set $E^\prime \subseteq E$. As a tree is a connected graph with no circuits, a spanning tree on $G$ is a spanning subgraph of $G$ that is a tree and hence $e(G') = v(G)-1$. The degree or coordination number $k_i$ of a vertex $v_i \in V$ is the number of edges attached to it.  A $k$-regular graph is a graph with the property that each of its vertices has the same degree $k$. In general, one can associate an edge weight $x_{ij}$ to each edge connecting adjacent vertices $v_i$ and $v_j$ (see, for example \cite{tzengwu}). For simplicity, all edge weights are set to one throughout this paper. A well-known method to enumerate spanning trees is to construct the Laplacian matrix $Q(G)$, which is the $v(G) \times v(G)$ matrix with element $Q(G)_{ij}=k_i\delta_{ij}-A(G)_{ij}$ where $A(G)$ is the adjacency matrix. One of the eigenvalues of $Q(G)$ is always zero. Denote the rest as $\lambda(G)_i$, $1 \le i \le v(G)-1$, then the number of spanning trees $N_{ST}(G) = (1/v(G))\prod_{i=1}^{v(G)-1} \lambda(G)_i$ \cite{bbook}.   

When the number of spanning trees $N_{ST}(G)$ grows exponentially with $v(G)$ as $v(G) \to \infty$, there exists a constant $z_G$ describing this exponential growth \cite{burton93,lyons05}:
\beq
z_G = \lim_{v(G) \to \infty} \frac{\ln N_{ST}(G)}{v(G)}
\label{zdef}
\eeq
where $G$, when used as a subscript in this manner, implicitly refers to
the thermodynamic limit.

For a $k$-regular graph $G_k$ with $k \ge 3$, there is a upper bound for the number of spanning trees \cite{mckay,chungyau}
\beq
N_{ST}(G_k) \le \Biggl ( \frac{2\ln v(G_k)}{v(G_k) k \ln k} \Bigg) (b_k)^{v(G_k)} \ ,
\label{nmckay}
\eeq
where
\beq
b_k = \frac{(k-1)^{k-1}}{[k(k-2)]^{\frac{k}{2}-1}} \ . 
\label{ck}
\eeq
With eq. (\ref{zdef}), this then yields \cite{sw} 
\beq
z_{G_k} \leq \ln(b_k) \ . 
\label{mcybound}
\eeq

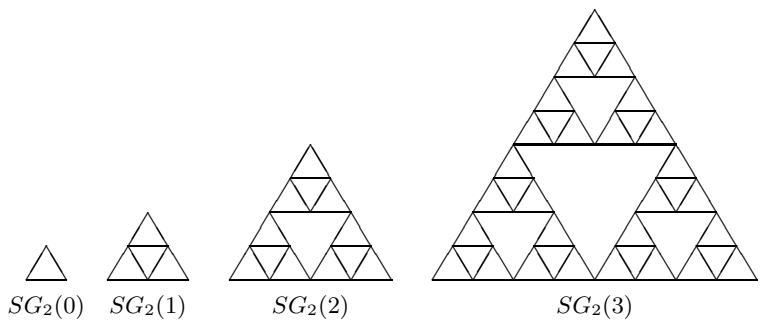
\begin{figure}[htbp]
\unitlength 0.9mm \hspace*{3mm}
\begin{picture}(108,40)
\put(0,0){\line(1,0){6}}
\put(0,0){\line(3,5){3}}
\put(6,0){\line(-3,5){3}}
\put(3,-4){\makebox(0,0){$SG_2(0)$}}
\put(12,0){\line(1,0){12}}
\put(12,0){\line(3,5){6}}
\put(24,0){\line(-3,5){6}}
\put(15,5){\line(1,0){6}}
\put(18,0){\line(3,5){3}}
\put(18,0){\line(-3,5){3}}
\put(18,-4){\makebox(0,0){$SG_2(1)$}}
\put(30,0){\line(1,0){24}}
\put(30,0){\line(3,5){12}}
\put(54,0){\line(-3,5){12}}
\put(36,10){\line(1,0){12}}
\put(42,0){\line(3,5){6}}
\put(42,0){\line(-3,5){6}}
\multiput(33,5)(12,0){2}{\line(1,0){6}}
\multiput(36,0)(12,0){2}{\line(3,5){3}}
\multiput(36,0)(12,0){2}{\line(-3,5){3}}
\put(39,15){\line(1,0){6}}
\put(42,10){\line(3,5){3}}
\put(42,10){\line(-3,5){3}}
\put(42,-4){\makebox(0,0){$SG_2(2)$}}
\put(60,0){\line(1,0){48}}
\put(72,20){\line(1,0){24}}
\put(60,0){\line(3,5){24}}
\put(84,0){\line(3,5){12}}
\put(84,0){\line(-3,5){12}}
\put(108,0){\line(-3,5){24}}
\put(66,10){\line(1,0){12}}
\put(90,10){\line(1,0){12}}
\put(78,30){\line(1,0){12}}
\put(72,0){\line(3,5){6}}
\put(96,0){\line(3,5){6}}
\put(84,20){\line(3,5){6}}
\put(72,0){\line(-3,5){6}}
\put(96,0){\line(-3,5){6}}
\put(84,20){\line(-3,5){6}}
\multiput(63,5)(12,0){4}{\line(1,0){6}}
\multiput(66,0)(12,0){4}{\line(3,5){3}}
\multiput(66,0)(12,0){4}{\line(-3,5){3}}
\multiput(69,15)(24,0){2}{\line(1,0){6}}
\multiput(72,10)(24,0){2}{\line(3,5){3}}
\multiput(72,10)(24,0){2}{\line(-3,5){3}}
\multiput(75,25)(12,0){2}{\line(1,0){6}}
\multiput(78,20)(12,0){2}{\line(3,5){3}}
\multiput(78,20)(12,0){2}{\line(-3,5){3}}
\put(81,35){\line(1,0){6}}
\put(84,30){\line(3,5){3}}
\put(84,30){\line(-3,5){3}}
\put(84,-4){\makebox(0,0){$SG_2(3)$}}
\end{picture}

\vspace*{5mm}
\caption{\footnotesize{The first four stages $n=0,1,2,3$ of the two-dimensional Sierpinski gasket $SG_2(n)$.}} 
\label{sgfig}
\end{figure}

The construction of the two-dimensional Sierpinski gasket $SG_2(n)$ at stage $n$ is shown in Fig. \ref{sgfig}. At stage $n=0$, it is an equilateral triangle; while stage $n+1$ is obtained by the juxtaposition of three $n$-stage structures. In general, the Sierpinski gaskets $SG_d$ can be built in any Euclidean dimension $d$ with fractal dimensionality $D=\ln(d+1)/\ln2$ \cite{Gefen81}. For the Sierpinski gasket $SG_d(n)$, the numbers of edges and vertices are given by 
\beq
e(SG_d(n)) = {d+1 \choose 2} (d+1)^n = \frac{d}{2} (d+1)^{n+1} \ ,
\label{e}
\eeq
\beq
v(SG_d(n)) = \frac{d+1}{2} [(d+1)^n+1] \ .
\label{v}
\eeq
Except the $(d+1)$ outmost vertices which have degree $d$, all other vertices of $SG_d(n)$ have degree $2d$. In the large $n$ limit, $SG_d$ is $2d$-regular. 

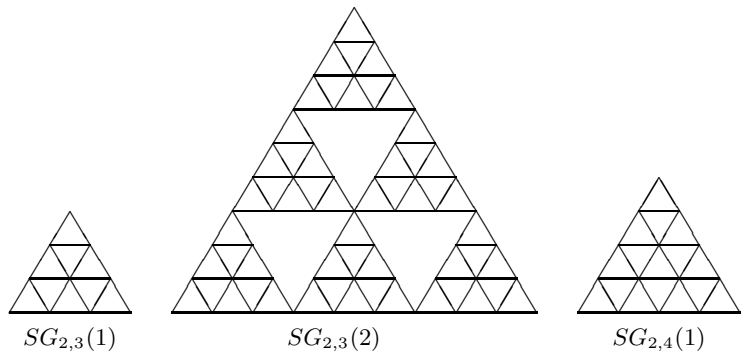
\begin{figure}[htbp]
\unitlength 0.9mm \hspace*{3mm}
\begin{picture}(108,45)
\put(0,0){\line(1,0){18}}
\put(3,5){\line(1,0){12}}
\put(6,10){\line(1,0){6}}
\put(0,0){\line(3,5){9}}
\put(6,0){\line(3,5){6}}
\put(12,0){\line(3,5){3}}
\put(18,0){\line(-3,5){9}}
\put(12,0){\line(-3,5){6}}
\put(6,0){\line(-3,5){3}}
\put(9,-4){\makebox(0,0){$SG_{2,3}(1)$}}
\put(24,0){\line(1,0){54}}
\put(33,15){\line(1,0){36}}
\put(42,30){\line(1,0){18}}
\put(24,0){\line(3,5){27}}
\put(42,0){\line(3,5){18}}
\put(60,0){\line(3,5){9}}
\put(78,0){\line(-3,5){27}}
\put(60,0){\line(-3,5){18}}
\put(42,0){\line(-3,5){9}}
\multiput(27,5)(18,0){3}{\line(1,0){12}}
\multiput(30,10)(18,0){3}{\line(1,0){6}}
\multiput(30,0)(18,0){3}{\line(3,5){6}}
\multiput(36,0)(18,0){3}{\line(3,5){3}}
\multiput(36,0)(18,0){3}{\line(-3,5){6}}
\multiput(30,0)(18,0){3}{\line(-3,5){3}}
\multiput(36,20)(18,0){2}{\line(1,0){12}}
\multiput(39,25)(18,0){2}{\line(1,0){6}}
\multiput(39,15)(18,0){2}{\line(3,5){6}}
\multiput(45,15)(18,0){2}{\line(3,5){3}}
\multiput(45,15)(18,0){2}{\line(-3,5){6}}
\multiput(39,15)(18,0){2}{\line(-3,5){3}}
\put(45,35){\line(1,0){12}}
\put(48,40){\line(1,0){6}}
\put(48,30){\line(3,5){6}}
\put(54,30){\line(3,5){3}}
\put(54,30){\line(-3,5){6}}
\put(48,30){\line(-3,5){3}}
\put(48,-4){\makebox(0,0){$SG_{2,3}(2)$}}
\put(84,0){\line(1,0){24}}
\put(87,5){\line(1,0){18}}
\put(90,10){\line(1,0){12}}
\put(93,15){\line(1,0){6}}
\put(84,0){\line(3,5){12}}
\put(90,0){\line(3,5){9}}
\put(96,0){\line(3,5){6}}
\put(102,0){\line(3,5){3}}
\put(108,0){\line(-3,5){12}}
\put(102,0){\line(-3,5){9}}
\put(96,0){\line(-3,5){6}}
\put(90,0){\line(-3,5){3}}
\put(96,-4){\makebox(0,0){$SG_{2,4}(1)$}}
\end{picture}

\vspace*{5mm}
\caption{\footnotesize{The generalized two-dimensional Sierpinski gasket $SG_{2,b}(n)$ with $b=3$ at stage $n=1, 2$ and $b=4$ at stage $n=1$.}} 
\label{sgbfig}
\end{figure}

The Sierpinski gasket can be generalized, denoted as $SG_{d,b}(n)$, by introducing the side length $b$ which is an integer larger or equal to two \cite{Hilfer}. The generalized Sierpinski gasket at stage $n+1$ is constructed with $b$ layers of stage $n$ hypertetrahedrons. The two-dimensional $SG_{2,b}(n)$ with $b=3$ at stage $n=1, 2$ and $b=4$ at stage $n=1$ are illustrated in Fig. \ref{sgbfig}. The ordinary Sierpinski gasket $SG_d(n)$ corresponds to the $b=2$ case, where the index $b$ is neglected for simplicity. The Hausdorff dimension for $SG_{d,b}$ is given by $D=\ln {b+d-1 \choose d} / \ln b$ \cite{Hilfer}. Notice that $SG_{d,b}$ is not $k$-regular even in the thermodynamic limit.

\section{The number of spanning trees on $SG_2(n)$}
\label{sectionIII}

In this section we derive the number of spanning trees on the two-dimensional Sierpinski gasket $SG_2(n)$ in detail. Let us start with the definitions of the quantities to be used.

\bigskip

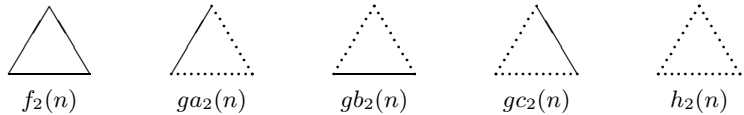
\begin{figure}[htbp]
\unitlength 1.8mm 
\begin{picture}(54,5)
\put(0,0){\line(1,0){6}}
\put(0,0){\line(3,5){3}}
\put(6,0){\line(-3,5){3}}
\put(3,-2){\makebox(0,0){$f_2(n)$}}
\multiput(12,0)(0.5,0){13}{\circle*{0.2}}
\put(12,0){\line(3,5){3}}
\multiput(18,0)(-0.3,0.5){11}{\circle*{0.2}}
\put(15,-2){\makebox(0,0){$ga_2(n)$}}
\put(24,0){\line(1,0){6}}
\multiput(24,0)(0.3,0.5){11}{\circle*{0.2}}
\multiput(30,0)(-0.3,0.5){11}{\circle*{0.2}}
\put(27,-2){\makebox(0,0){$gb_2(n)$}}
\multiput(36,0)(0.5,0){13}{\circle*{0.2}}
\multiput(36,0)(0.3,0.5){11}{\circle*{0.2}}
\put(42,0){\line(-3,5){3}}
\put(39,-2){\makebox(0,0){$gc_2(n)$}}
\multiput(48,0)(0.5,0){13}{\circle*{0.2}}
\multiput(48,0)(0.3,0.5){11}{\circle*{0.2}}
\multiput(54,0)(-0.3,0.5){11}{\circle*{0.2}}
\put(51,-2){\makebox(0,0){$h_2(n)$}}
\end{picture}

\vspace*{5mm}
\caption{\footnotesize{Illustration for the spanning subgraphs $f_2(n)$, $ga_2(n)$, $gb_2(n)$, $gc_2(n)$ and $h_2(n)$. The two outmost vertices at the ends of a solid line belong to one tree, while the two outmost vertices at the ends of a dot line belong to separated trees.}} 
\label{fghfig}
\end{figure}

\begin{defi} \label{defisg2} Consider the two-dimensional Sierpinski gasket $SG_2(n)$ at stage $n$. (a) Define $f_2(n) \equiv N_{ST}(SG_2(n))$ as the number of spanning trees. (b) Define $ga_2(n)$, $gb_2(n)$, $gc_2(n)$ as the number of spanning subgraphs with two trees such that one of the outmost vertices belongs to one tree and the other two outmost vertices belong to the other tree as illustrated in Fig. \ref{fghfig}. (c) Define $h_2(n)$ as the number of spanning subgraphs with three trees such that each of the outmost vertices belongs to a different tree.
\end{defi}

\bigskip

The quantities $f_2(n)$, $ga_2(n)$, $gb_2(n)$, $gc_2(n)$ and $h_2(n)$ are illustrated in Fig. \ref{fghfig}, where only the outmost vertices are shown. It is clear that the values $ga_2(n)$, $gb_2(n)$, $gc_2(n)$ are the same because of rotation symmetry, and we define $g_2(n) \equiv ga_2(n) = gb_2(n) = gc_2(n)$. The initial values at stage 0 are $f_2(0)=3$, $g_2(0)=1$, $h_2(0)=1$. Notice that $h_2(n)$ is the number of spanning trees on $SG_2(n)$ with the three outmost vertices identified.
The purpose is to obtain the expression of $f_2(n)$ as follows.

\bigskip

\begin{theo} \label{theosg2} The number of spanning trees on the two-dimensional Sierpinski gasket $SG_2(n)$ at stage $n$ is given by
\beq
f_2(n) = 2^{\alpha_2 (n)} 3^{\beta_2 (n)} 5^{\gamma_2 (n)} \ , 
\label{f}
\eeq
where the exponents are 
\beq
\alpha_2 (n) = \frac12 (3^n-1) \ , \
\label{alpha}
\eeq
\beq
\beta_2 (n) = \frac14 (3^{n+1}+2n+1) \ , 
\label{beta}
\eeq
\beq
\gamma_2 (n) = \frac14 (3^n-2n-1) \ .
\label{gamma}
\eeq
\end{theo}

\bigskip

This theorem can be proved by the following two lemmas.
The three quantities $f_2(n)$, $g_2(n)$ and $h_2(n)$ satisfy recursion relations which were first obtained in \cite{Dhar97}. 

\bigskip

\begin{lemma} \label{lemmasg2r} For any non-negative integer $n$,
\beq
f_2(n+1) = 6f_2^2(n)g_2(n) \ , 
\label{feq}
\eeq
\beq
g_2(n+1) = f_2^2(n)h_2(n) + 7 f_2(n)g_2^2(n) \ , 
\label{geq}
\eeq
\beq
h_2(n+1) = 12f_2(n)g_2(n)h_2(n) + 14g_2^3(n) \ .
\label{heq}
\eeq
\end{lemma}

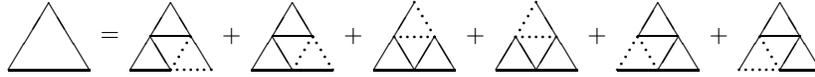
\begin{figure}[htbp]
\unitlength 0.9mm 
\begin{picture}(120,12)
\put(0,0){\line(1,0){12}}
\put(0,0){\line(3,5){6}}
\put(12,0){\line(-3,5){6}}
\put(15,5){\makebox(0,0){$=$}}
\put(18,0){\line(1,0){6}}
\put(21,5){\line(1,0){6}}
\put(18,0){\line(3,5){6}}
\put(30,0){\line(-3,5){6}}
\put(24,0){\line(-3,5){3}}
\multiput(24,0)(1,0){7}{\circle*{0.2}}
\multiput(24,0)(0.6,1){6}{\circle*{0.2}}
\put(33,5){\makebox(0,0){$+$}}
\put(36,0){\line(1,0){12}}
\put(39,5){\line(1,0){6}}
\put(36,0){\line(3,5){6}}
\multiput(42,0)(3,5){2}{\line(-3,5){3}}
\multiput(42,0)(0.6,1){6}{\circle*{0.2}}
\multiput(48,0)(-0.6,1){6}{\circle*{0.2}}
\put(51,5){\makebox(0,0){$+$}}
\put(54,0){\line(1,0){12}}
\put(54,0){\line(3,5){6}}
\put(60,0){\line(3,5){3}}
\multiput(60,0)(6,0){2}{\line(-3,5){3}}
\multiput(57,5)(1,0){7}{\circle*{0.2}}
\multiput(63,5)(-0.6,1){6}{\circle*{0.2}}
\put(69,5){\makebox(0,0){$+$}}
\put(72,0){\line(1,0){12}}
\put(84,0){\line(-3,5){6}}
\put(78,0){\line(-3,5){3}}
\multiput(72,0)(6,0){2}{\line(3,5){3}}
\multiput(75,5)(1,0){7}{\circle*{0.2}}
\multiput(75,5)(0.6,1){6}{\circle*{0.2}}
\put(87,5){\makebox(0,0){$+$}}
\put(90,0){\line(1,0){12}}
\put(102,0){\line(-3,5){6}}
\put(93,5){\line(1,0){6}}
\multiput(96,0)(-3,5){2}{\line(3,5){3}}
\multiput(90,0)(0.6,1){6}{\circle*{0.2}}
\multiput(96,0)(-0.6,1){6}{\circle*{0.2}}
\put(105,5){\makebox(0,0){$+$}}
\put(120,0){\line(-3,5){6}}
\put(108,0){\line(3,5){6}}
\put(114,0){\line(3,5){3}}
\multiput(114,0)(-3,5){2}{\line(1,0){6}}
\multiput(108,0)(1,0){7}{\circle*{0.2}}
\multiput(114,0)(-0.6,1){6}{\circle*{0.2}}
\end{picture}

\caption{\footnotesize{Illustration for the expression of  $f_2(n+1)$.}} 
\label{ffig}
\end{figure}

{\sl Proof} \quad 
The Sierpinski gaskets $SG_2(n+1)$ is composed of three $SG_2(n)$ with three pairs of vertices identified. To obtain the number of spanning trees $f_2(n+1)$, one of the $SG_2(n)$ should be spanned by two trees. There are six possibilities as illustrated in Fig. \ref{ffig}. Therefore, we have
\beq
f_2(n+1)=2f_2^2(n)[ga_2(n)+gb_2(n)+gc_2(n)]=6f_2^2(n)g_2(n) \ .
\eeq

\begin{figure}[htbp]
\unitlength 0.9mm 
\begin{picture}(84,12)
\put(0,0){\line(3,5){6}}
\multiput(0,0)(1,0){13}{\circle*{0.2}}
\multiput(12,0)(-0.6,1){11}{\circle*{0.2}}
\put(15,5){\makebox(0,0){$=$}}
\put(18,0){\line(3,5){6}}
\multiput(18,0)(3,5){2}{\line(1,0){6}}
\multiput(24,0)(3,5){2}{\line(-3,5){3}}
\multiput(24,0)(1,0){7}{\circle*{0.2}}
\multiput(24,0)(0.6,1){6}{\circle*{0.2}}
\multiput(30,0)(-0.6,1){6}{\circle*{0.2}}
\put(33,5){\makebox(0,0){$+$}}
\put(36,0){\line(3,5){6}}
\put(36,0){\line(1,0){6}}
\put(42,0){\line(3,5){3}}
\put(42,0){\line(-3,5){3}}
\multiput(42,0)(1,0){7}{\circle*{0.2}}
\multiput(39,5)(1,0){7}{\circle*{0.2}}
\multiput(48,0)(-0.6,1){11}{\circle*{0.2}}
\put(51,5){\makebox(0,0){$+$}}
\put(54,0){\line(3,5){6}}
\put(57,5){\line(1,0){6}}
\put(60,0){\line(3,5){3}}
\put(63,5){\line(-3,5){3}}
\multiput(54,0)(1,0){13}{\circle*{0.2}}
\multiput(60,0)(-0.6,1){6}{\circle*{0.2}}
\multiput(66,0)(-0.6,1){6}{\circle*{0.2}}
\put(69,5){\makebox(0,0){$+$}}
\put(72,0){\line(3,5){6}}
\put(78,0){\line(1,0){6}}
\put(78,0){\line(3,5){3}}
\put(84,0){\line(-3,5){3}}
\multiput(72,0)(1,0){7}{\circle*{0.2}}
\multiput(75,5)(1,0){7}{\circle*{0.2}}
\multiput(78,0)(-0.6,1){6}{\circle*{0.2}}
\multiput(81,5)(-0.6,1){6}{\circle*{0.2}}
\end{picture}

\unitlength 0.9mm 
\begin{picture}(84,12)
\put(15,5){\makebox(0,0){$+$}}
\put(18,0){\line(3,5){6}}
\put(18,0){\line(1,0){6}}
\multiput(24,0)(6,0){2}{\line(-3,5){3}}
\multiput(24,0)(1,0){7}{\circle*{0.2}}
\multiput(21,5)(1,0){7}{\circle*{0.2}}
\multiput(24,0)(0.6,1){6}{\circle*{0.2}}
\multiput(27,5)(-0.6,1){6}{\circle*{0.2}}
\put(33,5){\makebox(0,0){$+$}}
\put(36,0){\line(3,5){3}}
\put(36,0){\line(1,0){6}}
\put(42,0){\line(3,5){3}}
\multiput(42,0)(3,5){2}{\line(-3,5){3}}
\multiput(42,0)(1,0){7}{\circle*{0.2}}
\multiput(39,5)(1,0){7}{\circle*{0.2}}
\multiput(39,5)(0.6,1){6}{\circle*{0.2}}
\multiput(48,0)(-0.6,1){6}{\circle*{0.2}}
\put(51,5){\makebox(0,0){$+$}}
\put(54,0){\line(3,5){6}}
\put(63,5){\line(-3,5){3}}
\multiput(60,0)(-3,5){2}{\line(1,0){6}}
\multiput(54,0)(1,0){7}{\circle*{0.2}}
\multiput(60,0)(-0.6,1){6}{\circle*{0.2}}
\multiput(60,0)(0.6,1){6}{\circle*{0.2}}
\multiput(66,0)(-0.6,1){6}{\circle*{0.2}}
\put(69,5){\makebox(0,0){$+$}}
\multiput(72,0)(3,5){2}{\line(1,0){6}}
\multiput(78,0)(-3,5){2}{\line(3,5){3}}
\put(81,5){\line(-3,5){3}}
\multiput(78,0)(1,0){7}{\circle*{0.2}}
\multiput(72,0)(0.6,1){6}{\circle*{0.2}}
\multiput(78,0)(-0.6,1){6}{\circle*{0.2}}
\multiput(84,0)(-0.6,1){6}{\circle*{0.2}}
\end{picture}

\caption{\footnotesize{Illustration for the expression of  $ga_2(n+1)$.}} 
\label{gfig}
\end{figure}
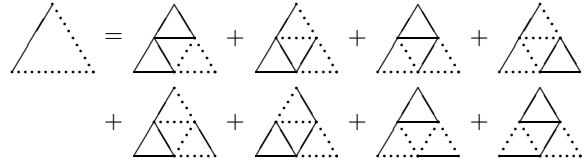

Similarly, $ga_2(n+1)$ for $SG_2(n+1)$ can be obtained with appropriated configurations of its three constituting $SG_2(n)$ as illustrated in Fig. \ref{gfig}. Thus,
\beqs
ga_2(n+1) & = & f_2^2(n)h_2(n) + 3f_2(n)ga_2^2(n) + 2f_2(n)ga_2(n)gc_2(n) \cr\cr
 & + & 2f_2(n)ga_2(n)gb_2(n) \ .
\eeqs
With the identity $ga_2(n)=gb_2(n)=gc_2(n)=g_2(n)$, eq. (\ref{geq}) is verified.

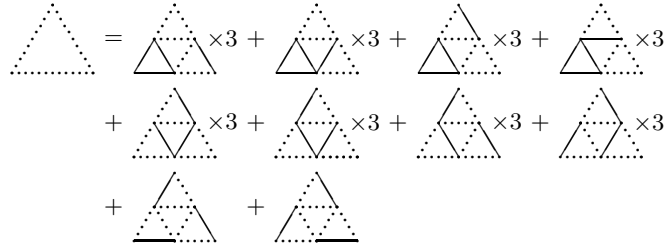
\begin{figure}[htbp]
\unitlength 0.9mm 
\begin{picture}(94,12)
\multiput(0,0)(1,0){13}{\circle*{0.2}}
\multiput(0,0)(0.6,1){11}{\circle*{0.2}}
\multiput(12,0)(-0.6,1){11}{\circle*{0.2}}
\put(15,5){\makebox(0,0){$=$}}
\put(18,0){\line(1,0){6}}
\put(18,0){\line(3,5){3}}
\multiput(24,0)(6,0){2}{\line(-3,5){3}}
\multiput(24,0)(1,0){7}{\circle*{0.2}}
\multiput(21,5)(1,0){7}{\circle*{0.2}}
\multiput(24,0)(0.6,1){6}{\circle*{0.2}}
\multiput(21,5)(0.6,1){6}{\circle*{0.2}}
\multiput(27,5)(-0.6,1){6}{\circle*{0.2}}
\put(31,5){\makebox(0,0){$\times 3$}}
\put(36,5){\makebox(0,0){$+$}}
\put(39,0){\line(1,0){6}}
\put(45,0){\line(-3,5){3}}
\multiput(39,0)(6,0){2}{\line(3,5){3}}
\multiput(45,0)(1,0){7}{\circle*{0.2}}
\multiput(42,5)(1,0){7}{\circle*{0.2}}
\multiput(42,5)(0.6,1){6}{\circle*{0.2}}
\multiput(51,0)(-0.6,1){11}{\circle*{0.2}}
\put(52,5){\makebox(0,0){$\times 3$}}
\put(57,5){\makebox(0,0){$+$}}
\put(60,0){\line(1,0){6}}
\put(60,0){\line(3,5){3}}
\multiput(66,0)(3,5){2}{\line(-3,5){3}}
\multiput(66,0)(1,0){7}{\circle*{0.2}}
\multiput(63,5)(1,0){7}{\circle*{0.2}}
\multiput(66,0)(0.6,1){6}{\circle*{0.2}}
\multiput(63,5)(0.6,1){6}{\circle*{0.2}}
\multiput(72,0)(-0.6,1){6}{\circle*{0.2}}
\put(73,5){\makebox(0,0){$\times 3$}}
\put(78,5){\makebox(0,0){$+$}}
\put(81,0){\line(3,5){3}}
\put(87,0){\line(-3,5){3}}
\multiput(81,0)(3,5){2}{\line(1,0){6}}
\multiput(87,0)(1,0){7}{\circle*{0.2}}
\multiput(87,0)(0.6,1){6}{\circle*{0.2}}
\multiput(84,5)(0.6,1){6}{\circle*{0.2}}
\multiput(93,0)(-0.6,1){11}{\circle*{0.2}}
\put(94,5){\makebox(0,0){$\times 3$}}
\end{picture}

\unitlength 0.9mm 
\begin{picture}(94,12)
\put(15,5){\makebox(0,0){$+$}}
\put(24,0){\line(3,5){3}}
\multiput(24,0)(3,5){2}{\line(-3,5){3}}
\multiput(18,0)(1,0){13}{\circle*{0.2}}
\multiput(18,0)(0.6,1){11}{\circle*{0.2}}
\multiput(21,5)(1,0){7}{\circle*{0.2}}
\multiput(30,0)(-0.6,1){6}{\circle*{0.2}}
\put(31,5){\makebox(0,0){$\times 3$}}
\put(36,5){\makebox(0,0){$+$}}
\put(45,0){\line(-3,5){3}}
\multiput(45,0)(-3,5){2}{\line(3,5){3}}
\multiput(39,0)(1,0){13}{\circle*{0.2}}
\multiput(51,0)(-0.6,1){11}{\circle*{0.2}}
\multiput(45,0)(1,0){7}{\circle*{0.2}}
\multiput(42,5)(1,0){7}{\circle*{0.2}}
\multiput(39,0)(0.6,1){6}{\circle*{0.2}}
\put(52,5){\makebox(0,0){$\times 3$}}
\put(57,5){\makebox(0,0){$+$}}
\put(63,5){\line(3,5){3}}
\multiput(66,0)(6,0){2}{\line(-3,5){3}}
\multiput(60,0)(1,0){13}{\circle*{0.2}}
\multiput(63,5)(1,0){7}{\circle*{0.2}}
\multiput(60,0)(0.6,1){6}{\circle*{0.2}}
\multiput(66,0)(0.6,1){6}{\circle*{0.2}}
\multiput(69,5)(-0.6,1){6}{\circle*{0.2}}
\put(73,5){\makebox(0,0){$\times 3$}}
\put(78,5){\makebox(0,0){$+$}}
\put(90,5){\line(-3,5){3}}
\multiput(81,0)(6,0){2}{\line(3,5){3}}
\multiput(81,0)(1,0){13}{\circle*{0.2}}
\multiput(84,5)(1,0){7}{\circle*{0.2}}
\multiput(84,5)(0.6,1){6}{\circle*{0.2}}
\multiput(87,0)(-0.6,1){6}{\circle*{0.2}}
\multiput(93,0)(-0.6,1){6}{\circle*{0.2}}
\put(94,5){\makebox(0,0){$\times 3$}}
\end{picture}

\unitlength 0.9mm 
\begin{picture}(94,12)
\put(15,5){\makebox(0,0){$+$}}
\put(18,0){\line(1,0){6}}
\put(21,5){\line(3,5){3}}
\put(30,0){\line(-3,5){3}}
\multiput(18,0)(0.6,1){6}{\circle*{0.2}}
\multiput(24,0)(0.6,1){6}{\circle*{0.2}}
\multiput(24,0)(1,0){7}{\circle*{0.2}}
\multiput(21,5)(1,0){7}{\circle*{0.2}}
\multiput(24,0)(-0.6,1){6}{\circle*{0.2}}
\multiput(27,5)(-0.6,1){6}{\circle*{0.2}}
\put(36,5){\makebox(0,0){$+$}}
\put(45,0){\line(1,0){6}}
\put(39,0){\line(3,5){3}}
\put(48,5){\line(-3,5){3}}
\multiput(39,0)(1,0){7}{\circle*{0.2}}
\multiput(42,5)(1,0){7}{\circle*{0.2}}
\multiput(42,5)(0.6,1){6}{\circle*{0.2}}
\multiput(45,0)(0.6,1){6}{\circle*{0.2}}
\multiput(45,0)(-0.6,1){6}{\circle*{0.2}}
\multiput(51,0)(-0.6,1){6}{\circle*{0.2}}
\end{picture}

\caption{\footnotesize{Illustration for the expression of  $h_2(n+1)$. The multiplication for the eight configurations on the right-hand-side corresponds to three possible orientations.}} 
\label{hfig}
\end{figure}

Finally, $h_2(n+1)$ is the summation of appropriated configurations as illustrated in Fig. \ref{hfig}, so that 
\beqs
& & h_2(n+1) \cr\cr
& = & 4f_2(n)h_2(n)[ga_2(n)+gb_2(n)+gc_2(n)] \cr\cr
& & + 2gc_2(n)ga_2(n)[gc_2(n)+ga_2(n)] + 2ga_2(n)gb_2(n)[ga_2(n)+gb_2(n)] \cr\cr
& & + 2gb_2(n)gc_2(n)[gb_2(n)+gc_2(n)] + 2ga_2(n)gb_2(n)gc_2(n) \ . \cr & & 
\eeqs
With the identity $ga_2(n)=gb_2(n)=gc_2(n)=g_2(n)$, eq. (\ref{heq}) is verified. \ $\Box$
 
\bigskip

By the three equations (\ref{feq}), (\ref{geq}) and (\ref{heq}), $f_2(n)$, $g_2(n)$ and $h_2(n)$ can be solved.

\bigskip

\begin{lemma} \label{lemmasg2s} For any non-negative integer $n$,
\beq
f_2(n) = 2^{\alpha_2 (n)} 3^{\beta_2 (n)} 5^{\gamma_2 (n)} \ , 
\label{fsol}
\eeq
\beq
g_2(n) = 2^{\alpha_2 (n)} 3^{\beta_2 (n)-n-1} 5^{\gamma_2 (n)+n} \ , 
\label{gsol}
\eeq
\beq
h_2(n) = 2^{\alpha_2 (n)} 3^{\beta_2 (n)-2n-1} 5^{\gamma_2 (n)+2n} \ .
\label{hsol}
\eeq
where $\alpha_2 (n)$, $\beta_2 (n)$ and $\gamma_2 (n)$ are given in eqs. (\ref{alpha})-(\ref{gamma})
\end{lemma}

{\sl Proof} \quad 
As $\alpha_2 (0)=0$, $\beta_2 (0)=1$ and $\gamma_2 (0)=0$, eqs. (\ref{fsol})-(\ref{hsol}) are correct for $n=0$. By the recursion relation (\ref{feq}), it is easy to obtain
\beq
f(n+1) = 2^{3\alpha_2 (n)+1} 3^{3\beta_2 (n)-n} 5^{3\gamma_2 (n)+n} 
\eeq
where 
\beq
3\alpha_2 (n)+1 = \frac12 (3^{n+1}-1) = \alpha_2 (n+1) \ , 
\eeq
\beq
3\beta_2 (n)-n = \frac14 (3^{n+2}+2n+3) = \beta_2 (n+1) \ , 
\eeq
\beq
3\gamma_2 (n)+n = \frac14 (3^{n+1}-2n-3) = \gamma_2 (n+1) \ .
\eeq
The proof for eq. (\ref{fsol}) is completed by induction. Similarly, we have
\beq
g(n+1) = 2^{3\alpha_2 (n)+1} 3^{3\beta_2 (n)-2n-2} 5^{3\gamma_2 (n)+2n+1} \ , 
\eeq
\beq
h(n+1) = 2^{3\alpha_2 (n)+1} 3^{3\beta_2 (n)-3n-3} 5^{3\gamma_2 (n)+3n+2} \ ,
\eeq
which verify eqs. (\ref{gsol}) and (\ref{hsol}).
\ $\Box$
 
\bigskip

Although the Laplacian matrix $Q(G)$ for $G=SG_2(n)$ can be constructed, it does not look simple to diagonalize. For example, with appropriate order of the vertices,
\beqs
Q(SG_2(1)) & = & \left( \begin{array}{cccccc}
4  & -1 & -1 & 0  & -1 & -1 \\
-1 & 4  & -1 & -1 & 0  & -1 \\
-1 & -1 & 4  & -1 & -1 & 0  \\
0  & -1 & -1 & 2  & 0  & 0  \\
-1 & 0  & -1 & 0  & 2  & 0  \\
-1 & -1 & 0  & 0  & 0  & 2
\end{array} \right )
\eeqs
and
\beqs
& & Q(SG_2(2)) \cr\cr
& = & \left( \begin{array}{ccccccccccccccc}
4  & -1 & -1 & 0  & 0  & 0  & 0  & 0  & 0  & 0  & -1 & -1 & 0  & 0  & 0  \\
-1 & 4  & -1 & 0  & 0  & 0  & 0  & 0  & 0  & 0  & -1 & 0  & -1 & 0  & 0  \\
-1 & -1 & 4  & 0  & 0  & 0  & 0  & 0  & 0  & 0  & 0  & -1 & -1 & 0  & 0  \\
0  & 0  & 0  & 4  & -1 & -1 & 0  & 0  & 0  & -1 & 0  & -1 & 0  & 0  & 0  \\
0  & 0  & 0  & -1 & 4  & -1 & 0  & 0  & 0  & 0  & 0  & -1 & 0  & 0  & -1 \\
0  & 0  & 0  & -1 & -1 & 4  & 0  & 0  & 0  & -1 & 0  & 0  & 0  & 0  & -1 \\
0  & 0  & 0  & 0  & 0  & 0  & 4  & -1 & -1 & -1 & -1 & 0  & 0  & 0  & 0  \\
0  & 0  & 0  & 0  & 0  & 0  & -1 & 4  & -1 & -1 & 0  & 0  & 0  & -1 & 0  \\
0  & 0  & 0  & 0  & 0  & 0  & -1 & -1 & 4  & 0  & -1 & 0  & 0  & -1 & 0  \\
0  & 0  & 0  & -1 & 0  & -1 & -1 & -1 & 0  & 4  & 0  & 0  & 0  & 0  & 0  \\
-1 & -1 & 0  & 0  & 0  & 0  & -1 & 0  & -1 & 0  & 4  & 0  & 0  & 0  & 0  \\
-1 & 0  & -1 & -1 & -1 & 0  & 0  & 0  & 0  & 0  & 0  & 4  & 0  & 0  & 0  \\
0  & -1 & -1 & 0  & 0  & 0  & 0  & 0  & 0  & 0  & 0  & 0  & 2  & 0  & 0  \\
0  & 0  & 0  & 0  & 0  & 0  & 0  & -1 & -1 & 0  & 0  & 0  & 0  & 2  & 0  \\
0  & 0  & 0  & 0  & -1 & -1 & 0  & 0  & 0  & 0  & 0  & 0  & 0  & 0  & 2
\end{array} \right ) \cr & &
\eeqs
Denote the non-zero eigenvalues of $Q(SG_2(n))$ as $\lambda(SG_2(n))_i$ for $1 \le i \le v(SG_2(n))-1$. 

\bigskip

\begin{cor} \label{corsg2} The product of non-zero eigenvalues of $Q(SG_2(n))$ is given by
\beq
\prod _{i=1}^{v(SG_2(n))-1} \lambda(SG_2(n))_i = \frac32 (3^n+1) f_2(n) \ ,
\eeq
where $f_2(n)$ is given in Theorem \ref{theosg2} and the number of vertices $v(SG_2(n))$ for $SG_2(n)$ is given by eq. (\ref{v}) with $d=2$.
\end{cor}

\bigskip

By the definition in eq. (\ref{zdef}), we have the following corollary.

\bigskip

\begin{cor} \label{corsg2a} The asymptotic growth constant for $SG_2$ is given by
\beqs
z_{SG_2} & = & \frac13 \ln 2 + \frac12 \ln 3 + \frac16 \ln 5 \cr\cr
& \simeq & 1.048594856...
\label{zsg2}
\eeqs
\end{cor}

\bigskip

This is equivalent to eq. (20) of \cite{Daerden}.
In passing, we notice that the number of spanning trees is the same for a planar graph and its dual, so that Theorem \ref{theosg2} also applies to the dual of $SG_2(n)$, denoted as $SG_2^*(n)$. As $SG_2$ is 4-regular in the large $n$ limit, we have $z_{SG_2^*}=z_{SG_2}$ \cite{sti}.

\section{The number of spanning trees on $SG_{2,b}(n)$ with $b=3,4$} 
\label{sectionIV}

The method given in the previous section can be applied to the number of spanning trees on $SG_{d,b}(n)$ with larger values of $d$ and $b$. The number of configurations to be considered increases as $d$ and $b$ increase, and the recursion relations must be derived individually for each $d$ and $b$. 
In this section, we consider the generalized two-dimensional Sierpinski gasket $SG_{2,b}(n)$ with the number of layers $b$ equal to three and four. 
For $SG_{2,3}(n)$, the numbers of edges and vertices are given by 
\beq
e(SG_{2,3}(n)) = 3 \times 6^n \ ,
\label{esg23}
\eeq
\beq
v(SG_{2,3}(n)) = \frac{7 \times 6^n + 8}{5} \ ,
\label{vsg23}
\eeq
where the three outmost vertices have degree two. There are $(6^n-1)/5$ vertices of $SG_{2,3}(n)$ with degree six and $6(6^n-1)/5$ vertices with degree four. Define $f_{2,3}(n)$, $g_{2,3}(n)$, $h_{2,3}(n)$ as in Definition \ref{defisg2} such that the initial values are $f_{2,3}(0)=3$, $g_{2,3}(0)=1$, $h_{2,3}(0)=1$. By the method illustrated in the above section, we obtain following recursion relations for any non-negative integer $n$.
\beq
f_{2,3}(n+1) = 18f_{2,3}^4(n)g_{2,3}(n)h_{2,3}(n) + 142f_{2,3}^3(n)g_{2,3}^3(n) \ , 
\label{f23eq}
\eeq
\beqs
g_{2,3}(n+1) & = & 2f_{2,3}^4(n)h_{2,3}^2(n) + 77 f_{2,3}^3(n)g_{2,3}^2(n)h_{2,3}(n) \cr\cr
 & + & 171f_{2,3}^2(n)g_{2,3}^4(n) \ , 
\label{g23eq}
\eeqs
\beqs
h_{2,3}(n+1) & = & 60f_{2,3}^3(n)g_{2,3}(n)h_{2,3}^2(n) + 564f_{2,3}^2(n)g_{2,3}^3(n)h_{2,3}(n) \cr\cr
& + & 468f_{2,3}(n)g_{2,3}^5(n) \ .
\label{h23eq}
\eeqs
The figures for these configurations are too many to be shown here.
By induction as in Lemma \ref{lemmasg2s}, these equations can be solved.

\bigskip

\begin{theo} \label{theosg23} The number of spanning trees on the two-dimensional Sierpinski gasket $SG_{2,3}(n)$ at stage $n$ is given by
\beq
f_{2,3}(n) = 2^{\alpha_{2,3} (n)} 3^{\beta_{2,3} (n)} 5^{\gamma_{2,3} (n)} 7^{\delta_{2,3} (n)} \ , 
\label{f23}
\eeq
where the exponents are 
\beq
\alpha_{2,3} (n) = \frac25 (6^n-1) \ ,
\label{alpha23}
\eeq
\beq
\beta_{2,3} (n) = \frac{1}{25} (13 \times 6^n-15n+12) \ ,
\label{beta23}
\eeq
\beq
\gamma_{2,3} (n) = \frac{1}{25} (3 \times 6^n-15n-3) \ ,
\label{gamma23}
\eeq
\beq
\delta_{2,3} (n) = \frac{1}{25} (7 \times 6^n+15n-7) \ .
\label{delta23}
\eeq
\end{theo}

\bigskip

$g_{2,3}(n)$ and $h_{2,3}(n)$ can also be expressed by these exponents:
\beqs
g_{2,3}(n) & = & 2^{\alpha_{2,3} (n)} 3^{\beta_{2,3} (n)+n-1} 5^{\gamma_{2,3} (n)+n} 7^{\delta_{2,3} (n)-n} \ , \cr\cr 
h_{2,3}(n) & = & 2^{\alpha_{2,3} (n)} 3^{\beta_{2,3} (n)+2n-1} 5^{\gamma_{2,3} (n)+2n} 7^{\delta_{2,3} (n)-2n} \ .
\eeqs

By the definition in eq. (\ref{zdef}), we have the following corollary.

\bigskip

\begin{cor} \label{corsg23a} The asymptotic growth constant for $SG_{2,3}$ is given by
\beqs
z_{SG_{2,3}} & = &  \frac27 \ln 2 + \frac{13}{35} \ln 3 + \frac{3}{35} \ln 5 + \frac15 \ln 7  \cr\cr
& \simeq & 1.133231895...
\label{zsg23}
\eeqs
\end{cor}

\bigskip

For $SG_{2,4}(n)$, the numbers of edges and vertices are given by 
\beq
e(SG_{2,4}(n)) = 3 \times 10^n \ ,
\label{esg24}
\eeq
\beq
v(SG_{2,4}(n)) = \frac{4 \times 10^n + 5}{3} \ ,
\label{vsg24}
\eeq
where the three outmost vertices have degree two. There are $(10^n-1)/3$ vertices of $SG_{2,4}(n)$ with degree six, and $(10^n-1)$ vertices with degree four. Define $f_{2,4}(n)$, $g_{2,4}(n)$, $h_{2,4}(n)$ as in Definition \ref{defisg2} such that the initial values are $f_{2,4}(0)=3$, $g_{2,4}(0)=1$, $h_{2,4}(0)=1$.
We write a computer program to obtain following recursion relations for any non-negative integer $n$.
\beqs
f_{2,4}(n+1) & = & 2f_{2,4}^7(n)h_{2,4}^3(n) + 516f_{2,4}^6(n)g_{2,4}^2(n)h_{2,4}^2(n) \cr\cr
 & + & 5856f_{2,4}^5(n)g_{2,4}^4(n)h_{2,4}(n) + 11354f_{2,4}^4(n)g_{2,4}^6(n) \ , 
\label{f24eq}
\eeqs
\beqs
g_{2,4}(n+1) & = & 82f_{2,4}^6(n)g_{2,4}(n)h_{2,4}^3(n) + 2786f_{2,4}^5(n)g_{2,4}^3(n)h_{2,4}^2(n) \cr\cr
 & + & 14480f_{2,4}^4(n)g_{2,4}^5(n)h_{2,4}(n) + 13732f_{2,4}^3(n)g_{2,4}^7(n) \ , 
\label{g24eq}
\eeqs
\beqs
h_{2,4}(n+1) & = & 20f_{2,4}^6(n)h_{2,4}^4(n) + 2388f_{2,4}^5(n)g_{2,4}^2(n)h_{2,4}^3(n) \cr\cr
& + & 30948f_{2,4}^4(n)g_{2,4}^4(n)h_{2,4}^2(n) + 83234f_{2,4}^3(n)g_{2,4}^6(n)h_{2,4}(n) \cr\cr
 & + & 42210f_{2,4}^2(n)g_{2,4}^8(n) \ .
\label{h24eq}
\eeqs
By induction as in Lemma \ref{lemmasg2s}, these equations can be solved.

\bigskip

\begin{theo} \label{theosg24} The number of spanning trees on the two-dimensional Sierpinski gasket $SG_{2,4}(n)$ at stage $n$ is given by
\beq
f_{2,4}(n) = 2^{\alpha_{2,4} (n)} 3^{\beta_{2,4} (n)} 5^{\gamma_{2,4} (n)} 41^{\delta_{2,4} (n)} 103^{\epsilon_{2,4} (n)} \ , 
\label{f24}
\eeq
where the exponents are 
\beq
\alpha_{2,4} (n) = \frac29 (10^n-1) \ ,
\label{alpha24}
\eeq
\beq
\beta_{2,4} (n) = \frac13 (10^n+2) \ ,
\label{beta24}
\eeq
\beq
\gamma_{2,4} (n) = \frac19 (10^n-1) \ ,
\label{gamma24}
\eeq
\beq
\delta_{2,4} (n) = \frac{2}{27} (2 \times 10^n+9n-2) \ ,
\label{delta24}
\eeq
\beq
\epsilon_{2,4} (n) = \frac{2}{27} (10^n-9n-1) \ .
\label{epsilon24}
\eeq
\end{theo}

\bigskip

$g_{2,4}(n)$ and $h_{2,4}(n)$ can also be expressed by these exponents:
\beqs
g_{2,4}(n) & = & 2^{\alpha_{2,4} (n)} 3^{\beta_{2,4} (n)-1} 5^{\gamma_{2,4} (n)} 41^{\delta_{2,4} (n)-n} 103^{\epsilon_{2,4} (n)+n} \ , \cr\cr 
h_{2,4}(n) & = & 2^{\alpha_{2,4} (n)} 3^{\beta_{2,4} (n)-1} 5^{\gamma_{2,4} (n)} 41^{\delta_{2,4} (n)-2n} 103^{\epsilon_{2,4} (n)+2n} \ .
\eeqs

By the definition in eq. (\ref{zdef}), we have the following corollary.

\bigskip

\begin{cor} \label{corsg24a} The asymptotic growth constant for $SG_{2,4}$ is given by
\beqs
z_{SG_{2,4}} & = &  \frac16 \ln 2 + \frac14 \ln 3 + \frac{1}{12} \ln 5 + \frac19 \ln (41) + \frac{1}{18} \ln (103) \cr\cr
& \simeq & 1.194401490...
\label{zsg24}
\eeqs
\end{cor}

\bigskip

\section{The number of spanning trees on $SG_d(n)$ with $d=3,4$} 
\label{sectionV}

In this section, we derive the number of spanning trees on $SG_d(n)$ with $d=3,4$.
For the three-dimensional Sierpinski gasket $SG_3(n)$, we use the following definitions.

\bigskip

\begin{defi} \label{defisg3} Consider the three-dimensional Sierpinski gasket $SG_3(n)$ at stage $n$. (a) Define $f_3(n) \equiv N_{ST}(SG_3(n))$ as the number of spanning trees. (b) Define $g_3(n)$ as the number of spanning subgraphs with two trees such that one of the outmost vertices belongs to one tree and the other three outmost vertices belong to the other tree. (c) Define $h_3(n)$ as the number of spanning subgraphs with two trees such that two of the outmost vertices belong to one tree and the other two outmost vertices belong to the other tree. (d) Define $p_3(n)$ as the number of spanning subgraphs with three trees such that two of the outmost vertices belong to one tree and the other two outmost vertices separately belong to the other trees. (e) Define $q_3(n)$ as the number of spanning subgraphs with four trees such that each of the outmost vertices belongs to a different tree.
\end{defi}

\bigskip

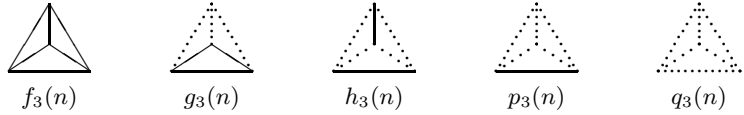
\begin{figure}[htbp]
\unitlength 1.8mm 
\begin{picture}(54,5)
\put(0,0){\line(1,0){6}}
\put(0,0){\line(3,5){3}}
\put(6,0){\line(-3,5){3}}
\put(0,0){\line(3,2){3}}
\put(6,0){\line(-3,2){3}}
\put(3,2){\line(0,1){3}}
\put(3,-2){\makebox(0,0){$f_3(n)$}}
\put(12,0){\line(1,0){6}}
\multiput(12,0)(0.3,0.5){11}{\circle*{0.2}}
\multiput(18,0)(-0.3,0.5){11}{\circle*{0.2}}
\multiput(15,2)(0,0.5){7}{\circle*{0.2}}
\put(12,0){\line(3,2){3}}
\put(18,0){\line(-3,2){3}}
\put(15,-2){\makebox(0,0){$g_3(n)$}}
\put(24,0){\line(1,0){6}}
\multiput(24,0)(0.3,0.5){11}{\circle*{0.2}}
\multiput(30,0)(-0.3,0.5){11}{\circle*{0.2}}
\put(27,2){\line(0,1){3}}
\multiput(24,0)(0.6,0.4){6}{\circle*{0.2}}
\multiput(30,0)(-0.6,0.4){6}{\circle*{0.2}}
\put(27,-2){\makebox(0,0){$h_3(n)$}}
\put(36,0){\line(1,0){6}}
\multiput(36,0)(0.3,0.5){11}{\circle*{0.2}}
\multiput(42,0)(-0.3,0.5){11}{\circle*{0.2}}
\multiput(39,2)(0,0.5){7}{\circle*{0.2}}
\multiput(36,0)(0.6,0.4){6}{\circle*{0.2}}
\multiput(42,0)(-0.6,0.4){6}{\circle*{0.2}}
\put(39,-2){\makebox(0,0){$p_3(n)$}}
\multiput(48,0)(0.5,0){13}{\circle*{0.2}}
\multiput(48,0)(0.3,0.5){11}{\circle*{0.2}}
\multiput(54,0)(-0.3,0.5){11}{\circle*{0.2}}
\multiput(51,2)(0,0.5){7}{\circle*{0.2}}
\multiput(48,0)(0.6,0.4){6}{\circle*{0.2}}
\multiput(54,0)(-0.6,0.4){6}{\circle*{0.2}}
\put(51,-2){\makebox(0,0){$q_3(n)$}}
\end{picture}

\vspace*{5mm}
\caption{\footnotesize{Illustration for the spanning subgraphs $f_3(n)$, $g_3(n)$, $h_3(n)$, $p_3(n)$ and $q_3(n)$. The two outmost vertices at the ends of a solid line belong to one tree, while the two outmost vertices at the ends of a dot line belong to separated trees.}} 
\label{fghpqfig}
\end{figure}

The quantities $f_3(n)$, $g_3(n)$, $h_3(n)$, $p_3(n)$ and $q_3(n)$ are illustrated in Fig. \ref{fghpqfig}, where only the outmost vertices are shown. There are four equivalent $g_3(n)$, three equivalent $h_3(n)$, and six equivalent $p_3(n)$. The initial values at stage 0 are $f_3(0)=16$, $g_3(0)=3$, $h_3(0)=1$, $p_3(0)=1$, $q_3(0)=1$. Notice that $q_3(n)$ is the number of spanning trees on $SG_3(n)$ with the four outmost vertices identified.

In the process of calculation, we find that it is convenient to combine $g_3(n)$ and $h_3(n)$ and define $gh_3(n) \equiv g_3(n)+h_3(n)$. We obtain following recursion relations for any non-negative integer $n$.
\beq
f_3(n+1) = 72f_3^2(n)gh_3(n)p_3(n) + 56f_3(n)gh_3^3(n) \ , 
\label{f3eq}
\eeq
\beqs
gh_3(n+1) & = & 6f_3^2(n)gh_3(n)q_3(n) + 26 f_3^2(n)p_3^2(n) \cr\cr
 & + & 120f_3(n)gh_3^2(n)p_3(n) + 22gh_3^4(n) \ , 
\label{gh3eq}
\eeqs
\beqs
p_3(n+1) & = & 6f_3^2(n)p_3(n)q_3(n) + 14f_3(n)gh_3^2(n)q_3(n) \cr\cr
& + & 120f_3(n)gh_3(n)p_3^2(n) + 88gh_3^3(n)p_3(n) \ ,
\label{p3eq}
\eeqs
\beqs
q_3(n+1) & = & 144f_3(n)gh_3(n)p_3(n)q_3(n) + 208f_3(n)p_3^3(n) \cr\cr
& + & 56gh_3^3(n)q_3(n) + 720gh_3^2(n)p_3^2(n) \ .
\label{q3eq}
\eeqs
The figures for these configurations are too many to be shown here.
With the initial values given above, these equations can be solved by induction as in Lemma \ref{lemmasg2s}.

\bigskip

\begin{theo} \label{theosg3} The number of spanning trees on the three-dimensional Sierpinski gasket $SG_3(n)$ at stage $n$ is given by
\beq
f_3(n) = 2^{\alpha_3 (n)} 3^{\beta_3 (n)} \ , 
\label{f3}
\eeq
where the exponents are 
\beq
\alpha_3 (n) = 4^{n+1}+n \ ,
\label{alpha3}
\eeq
\beq
\beta_3 (n) = \frac13 (4^n-3n-1) \ .
\label{beta3}
\eeq
\end{theo}

\bigskip

$gh_3(n)$, $p_3(n)$ and $q_3(n)$ can also be expressed by these exponents:
\beqs
gh_3(n) & = & 2^{\alpha_3 (n)-n-2} 3^{\beta_3 (n)+n} \ , \cr\cr 
p_3(n) & = & 2^{\alpha_3 (n)-2n-4} 3^{\beta_3 (n)+2n} \ , \cr\cr 
q_3(n) & = & 2^{\alpha_3 (n)-3n-4} 3^{\beta_3 (n)+3n} \ .
\eeqs

By the definition in eq. (\ref{zdef}), we have the following corollary.

\bigskip

\begin{cor} \label{corsg3a} The asymptotic growth constant for $SG_3$ is given by
\beqs
z_{SG_3} & = & 2 \ln 2 + \frac16 \ln 3 \cr\cr
& \simeq & 1.569396409...
\label{zsg3}
\eeqs
\end{cor}

\bigskip

For the four-dimensional Sierpinski gasket $SG_4(n)$, we use the following definitions.

\bigskip

\begin{defi} \label{defisg4} Consider the four-dimensional Sierpinski gasket $SG_4(n)$ at stage $n$. (a) Define $f_4(n) \equiv N_{ST}(SG_4(n))$ as the number of spanning trees. (b) Define $g_4(n)$ as the number of spanning subgraphs with two trees such that two of the outmost vertices belong to one tree and the other three outmost vertices belong to the other tree. (c) Define $h_4(n)$ as the number of spanning subgraphs with two trees such that one of the outmost vertices belong to one tree and the other four outmost vertices belong to the other tree. (d) Define $p_4(n)$ as the number of spanning subgraphs with three trees such that one of the outmost vertices belong to one tree, two of the other outmost vertices belong to another tree and the rest two outmost vertices belong to the other tree. (e) Define $q_4(n)$ as the number of spanning subgraphs with three trees such that three of the outmost vertices belong to one tree and the other two outmost vertices separately belong to the other trees. (f) Define $r_4(n)$ as the number of spanning subgraphs with four trees such that two of the outmost vertices belong to one tree and the other three outmost vertices separately belong to the other trees. (g) Define $s_4(n)$ as the number of spanning subgraphs with five trees such that each of the outmost vertices belongs to a different tree.
\end{defi}

\bigskip

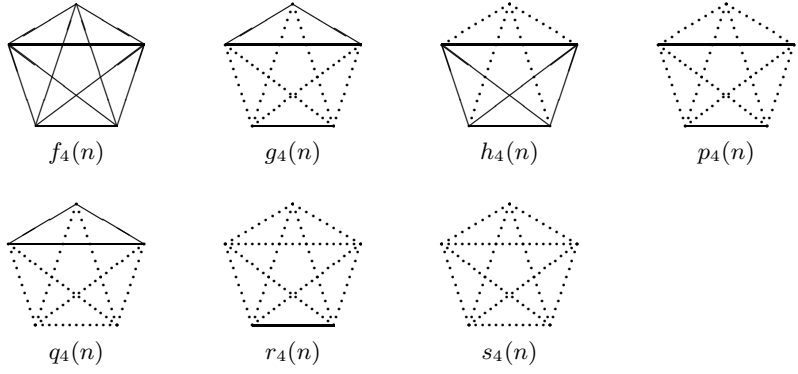
\begin{figure}[htbp]
\unitlength 1.8mm 
\begin{picture}(58,9)
\put(2,0){\line(1,0){6}}
\put(2,0){\line(4,3){8}}
\put(2,0){\line(-1,3){2}}
\put(2,0){\line(1,3){3}}
\put(8,0){\line(-1,3){3}}
\put(8,0){\line(1,3){2}}
\put(8,0){\line(-4,3){8}}
\put(0,6){\line(1,0){10}}
\put(5,9){\line(5,-3){5}}
\put(5,9){\line(-5,-3){5}}
\put(5,-2){\makebox(0,0){$f_4(n)$}}
\put(18,0){\line(1,0){6}}
\multiput(18,0)(0.4,0.3){21}{\circle*{0.2}}
\multiput(18,0)(-0.2,0.6){11}{\circle*{0.2}}
\multiput(18,0)(0.2,0.6){16}{\circle*{0.2}}
\multiput(24,0)(-0.2,0.6){16}{\circle*{0.2}}
\multiput(24,0)(0.2,0.6){11}{\circle*{0.2}}
\multiput(24,0)(-0.4,0.3){21}{\circle*{0.2}}
\put(16,6){\line(1,0){10}}
\put(21,9){\line(5,-3){5}}
\put(21,9){\line(-5,-3){5}}
\put(21,-2){\makebox(0,0){$g_4(n)$}}
\put(34,0){\line(1,0){6}}
\put(34,0){\line(4,3){8}}
\put(34,0){\line(-1,3){2}}
\multiput(34,0)(0.2,0.6){16}{\circle*{0.2}}
\multiput(40,0)(-0.2,0.6){16}{\circle*{0.2}}
\put(40,0){\line(1,3){2}}
\put(40,0){\line(-4,3){8}}
\put(32,6){\line(1,0){10}}
\multiput(37,9)(0.5,-0.3){11}{\circle*{0.2}}
\multiput(37,9)(-0.5,-0.3){11}{\circle*{0.2}}
\put(37,-2){\makebox(0,0){$h_4(n)$}}
\put(50,0){\line(1,0){6}}
\multiput(50,0)(0.4,0.3){21}{\circle*{0.2}}
\multiput(50,0)(-0.2,0.6){11}{\circle*{0.2}}
\multiput(50,0)(0.2,0.6){16}{\circle*{0.2}}
\multiput(56,0)(-0.2,0.6){16}{\circle*{0.2}}
\multiput(56,0)(0.2,0.6){11}{\circle*{0.2}}
\multiput(56,0)(-0.4,0.3){21}{\circle*{0.2}}
\put(48,6){\line(1,0){10}}
\multiput(53,9)(0.5,-0.3){11}{\circle*{0.2}}
\multiput(53,9)(-0.5,-0.3){11}{\circle*{0.2}}
\put(53,-2){\makebox(0,0){$p_4(n)$}}
\end{picture}

\vspace*{10mm}

\begin{picture}(58,9)
\multiput(2,0)(0.5,0){13}{\circle*{0.2}}
\multiput(2,0)(0.4,0.3){21}{\circle*{0.2}}
\multiput(2,0)(-0.2,0.6){11}{\circle*{0.2}}
\multiput(2,0)(0.2,0.6){16}{\circle*{0.2}}
\multiput(8,0)(-0.2,0.6){16}{\circle*{0.2}}
\multiput(8,0)(0.2,0.6){11}{\circle*{0.2}}
\multiput(8,0)(-0.4,0.3){21}{\circle*{0.2}}
\put(0,6){\line(1,0){10}}
\put(5,9){\line(5,-3){5}}
\put(5,9){\line(-5,-3){5}}
\put(5,-2){\makebox(0,0){$q_4(n)$}}
\put(18,0){\line(1,0){6}}
\multiput(18,0)(0.4,0.3){21}{\circle*{0.2}}
\multiput(18,0)(-0.2,0.6){11}{\circle*{0.2}}
\multiput(18,0)(0.2,0.6){16}{\circle*{0.2}}
\multiput(24,0)(-0.2,0.6){16}{\circle*{0.2}}
\multiput(24,0)(0.2,0.6){11}{\circle*{0.2}}
\multiput(24,0)(-0.4,0.3){21}{\circle*{0.2}}
\multiput(16,6)(0.5,0){21}{\circle*{0.2}}
\multiput(21,9)(0.5,-0.3){11}{\circle*{0.2}}
\multiput(21,9)(-0.5,-0.3){11}{\circle*{0.2}}
\put(21,-2){\makebox(0,0){$r_4(n)$}}
\multiput(34,0)(0.5,0){13}{\circle*{0.2}}
\multiput(34,0)(0.4,0.3){21}{\circle*{0.2}}
\multiput(34,0)(-0.2,0.6){11}{\circle*{0.2}}
\multiput(34,0)(0.2,0.6){16}{\circle*{0.2}}
\multiput(40,0)(-0.2,0.6){16}{\circle*{0.2}}
\multiput(40,0)(0.2,0.6){11}{\circle*{0.2}}
\multiput(40,0)(-0.4,0.3){21}{\circle*{0.2}}
\multiput(32,6)(0.5,0){21}{\circle*{0.2}}
\multiput(37,9)(0.5,-0.3){11}{\circle*{0.2}}
\multiput(37,9)(-0.5,-0.3){11}{\circle*{0.2}}
\put(37,-2){\makebox(0,0){$s_4(n)$}}
\end{picture}

\vspace*{5mm}
\caption{\footnotesize{Illustration for the spanning subgraphs $f_4(n)$, $g_4(n)$, $h_4(n)$, $p_4(n)$, $q_4(n)$, $r_4(n)$ and $s_4(n)$. The two outmost vertices at the ends of a solid line belong to one tree, while the two outmost vertices at the ends of a dot line belong to separated trees.}} 
\label{fghpqrsfig}
\end{figure}

The quantities $f_4(n)$, $g_4(n)$, $h_4(n)$, $p_4(n)$, $q_4(n)$, $r_4(n)$ and $s_4(n)$ are illustrated in Fig. \ref{fghpqrsfig}, where only the outmost vertices are shown. There are ten equivalent $g_4(n)$, five equivalent $h_4(n)$, fifteen equivalent $p_3(n)$, ten equivalent $q_4(n)$ and ten equivalent $r_4(n)$. The initial values at stage 0 are $f_4(0)=125$, $g_4(0)=3$, $h_4(0)=16$, $p_4(0)=1$, $q_4(0)=3$, $r_4(0)=1$, $s_4(0)=1$. Notice that $s_4(n)$ is the number of spanning trees on $SG_4(n)$ with the five outmost vertices identified.

We find that it is convenient to reduce the number of variables by defining $gh_4(n) \equiv 3g_4(n)+h_4(n)$ and $pq_4(n) \equiv 2p_4(n)+q_4(n)$. We write a computer program to obtain following recursion relations for any non-negative integer $n$.
\beqs
f_4(n+1) & = & 1440f_4^2(n)gh_4(n)pq_4(n)r_4(n) + 520f_4^2(n)pq_4^3(n) \cr\cr
 & + & 1120f_4(n)gh_4^3(n)r(n) + 3600f_4(n)gh_4^2(n)pq_4^2(n) \cr\cr
 & + & 1320gh_4^4(n)pq_4(n) \ , 
\label{f4eq}
\eeqs
\beqs
gh_4(n+1) & = & 72f_4^2(n)gh_4(n)pq_4(n)s_4(n) + 378f_4^2(n)gh_4(n)r_4^2(n) \cr\cr
 & + & 816 f_4^2(n)pq_4^2(n)r_4(n) + 56f_4(n)gh_4^3(n)s_4(n) \cr\cr
 & + & 3756f_4(n)gh_4^2(n)pq_4(n)r_4(n) + 2360f_4(n)gh_4(n)pq_4^3(n) \cr\cr
 & + & 688gh_4^4(n)r_4(n) + 2562gh_4^3(n)pq_4^2(n) \ , 
\label{gh4eq}
\eeqs
\beqs
pq_4(n+1) & = & 48f_4^2(n)gh_4(n)r_4(n)s_4(n) + 52f_4^2(n)pq_4^2(n)s_4(n) \cr\cr
 & + & 544f_4^2(n)pq_4(n)r_4^2(n) + 240f_4(n)gh_4^2(n)pq_4(n)s_4(n) \cr\cr
 & + & 1252f_4(n)gh_4^2(n)r_4^2(n) + 4720f_4(n)gh_4(n)pq_4^2(n)r_4(n) \cr\cr
 & + & 724f_4(n)pq_4^4(n) + 44gh_4^4(n)s_4(n) \cr\cr
 & + & 3416gh_4^3(n)pq_4(n)r_4(n) + 3104gh_4^2(n)pq_4^3(n) \ ,
\label{pq4eq}
\eeqs
\beqs
r_4(n+1) & = & 72f_4^2(n)pq_4(n)r_4(n)s_4(n) + 126f_4^2(n)r_4^3(n) \cr\cr
 & + & 168f_4(n)gh_4^2(n)r_4(n)s_4(n) + 360f_4(n)gh_4(n)pq_4^2(n)s_4(n) \cr\cr
 & + & 3756f_4(n)gh_4(n)pq_4(n)r_4^2(n) + 2360f_4(n)pq_4^3(n)r_4(n) \cr\cr
 & + & 264gh_4^3(n)pq_4(n)s_4(n) + 1376gh_4^3(n)r_4^2(n) \cr\cr
 & + & 7686gh_4^2(n)pq_4^2(n)r_4(n) + 2328gh_4(n)pq_4^4(n) \ ,
\label{r4eq}
\eeqs
\beqs
s_4(n+1) & = & 2880f_4(n)gh_4(n)pq_4(n)r_4(n)s_4(n) + 5040f_4(n)gh_4(n)r_4^3(n) \cr\cr
 & + & 1040f_4(n)pq_4^3(n)s_4(n) + 16320f_4(n)pq_4^2(n)r_4^2(n) \cr\cr
 & + & 1120gh_4^3(n)r_4(n)s_4(n) + 3600gh_4^2(n)pq_4^2(n)s_4(n) \cr\cr
 & + & 37560gh_4^2(n)pq_4(n)r_4^2(n) + 47200gh_4(n)pq_4^3(n)r_4(n) \cr\cr
 & + & 4344pq_4^5(n) \ .
\label{s4eq}
\eeqs

With the initial values given above, these equations can be solved by induction as in Lemma \ref{lemmasg2s}.

\bigskip

\begin{theo} \label{theosg4} The number of spanning trees on the four-dimensional Sierpinski gasket $SG_4(n)$ at stage $n$ is given by
\beq
f_4(n) = 2^{\alpha_4 (n)} 5^{\beta_4 (n)} 7^{\gamma_4 (n)} \ , 
\label{f4}
\eeq
where the exponents are 
\beq
\alpha_4 (n) = \frac32 (5^n-1) \ ,
\label{alpha4}
\eeq
\beq
\beta_4 (n) = \frac38 (5^{n+1}+4n+3) \ ,
\label{beta4}
\eeq
\beq
\gamma_4 (n) = \frac38 (5^n-4n-1) \ .
\label{gamma4}
\eeq
\end{theo}

\bigskip

$gh_4(n)$, $pq_4(n)$, $r_4(n)$ and $s_4(n)$ can also be expressed by these exponents:
\beqs
gh_4(n) & = & 2^{\alpha_4 (n)} 5^{\beta_4 (n)-n-1} 7^{\gamma_4 (n)+n} \ , \cr\cr 
pq_4(n) & = & 2^{\alpha_4 (n)} 5^{\beta_4 (n)-2n-2} 7^{\gamma_4 (n)+2n} \ , \cr\cr 
r_4(n) & = & 2^{\alpha_4 (n)} 5^{\beta_4 (n)-3n-3} 7^{\gamma_4 (n)+3n} \ , \cr\cr 
s_4(n) & = & 2^{\alpha_4 (n)} 5^{\beta_4 (n)-4n-3} 7^{\gamma_4 (n)+4n} \ .
\eeqs

By the definition in eq. (\ref{zdef}), we have the following corollary.

\bigskip

\begin{cor} \label{corsg4a} The asymptotic growth constant for $SG_4$ is given by
\beqs
z_{SG_4} & = & \frac35 \ln 2 + \frac34 \ln 5 + \frac{3}{20} \ln 7 \cr\cr
& \simeq & 1.914853265...
\label{zsg4}
\eeqs
\end{cor}

\bigskip

\section{The number of spanning trees on $SG_d(n)$ for general $d$} 
\label{sectionVI}

With the above theorems that the numbers of spanning trees on $SG_2(n)$, $SG_3(n)$ and $SG_4(n)$ have only simple factors, we observe the following conjecture for the number of spanning trees $f_d(n) \equiv N_{ST}(SG_d(n))$ on $SG_d(n)$ for general $d$.

\bigskip

\begin{conj} \label{conjsgd} The number of spanning trees on the $d$-dimensional Sierpinski gasket $SG_d(n)$ at stage $n$ is given by
\beq
f_d(n) = 2^{\alpha_d (n)} (d+1)^{\beta_d (n)} (d+3)^{\gamma_d (n)} \ , 
\label{fdsol}
\eeq
where the exponents are 
\beq
\alpha_d (n) = \frac{d-1}{2} [(d+1)^n-1] \ , \
\label{alphad}
\eeq
\beq
\beta_d (n) = \frac{d-1}{2d} [(d+1)^{n+1}+dn+d-1] \ , 
\label{betad}
\eeq
\beq
\gamma_d (n) = \frac{d-1}{2d} [(d+1)^n-dn-1] \ .
\label{gammad}
\eeq
\end{conj}

\bigskip

Notice that these exponents are positive integers when $d$ is a positive integer and $n$ is a non-negative integer. It is interesting that the recursion relations (\ref{f23eq})-(\ref{h23eq}), (\ref{f24eq})-(\ref{h24eq}), (\ref{f3eq})-(\ref{q3eq}) and (\ref{f4eq})-(\ref{s4eq}) are very complicated but the solutions are very simple. We do not know how to explain it. There may be a better method to solve this problem. 

$SG_d(0)$ at state $n=0$ is a complete graph with $(d+1)$ vertices, each of which is adjacent to all of the other vertices. Setting $n=0$ into Conjecture \ref{conjsgd} is consistent with the expectation that the number of spanning trees on the $d$-dimensional Sierpinski gasket $SG_d(0)$ at stage zero is given by $f_d(0) = (d+1)^{d-1}$.

With the number of vertices for $SG_d(n)$ given in eq. (\ref{v}), Conjecture \ref{conjsgd} leads to following result.

\bigskip

\begin{conj} \label{conjsgda} The asymptotic growth constant for $SG_d(n)$ is given by
\beq
z_{SG_d} = \frac{d-1}{d(d+1)} [ d \ln 2 + (d+1) \ln (d+1) + \ln (d+3) ] \ .
\label{zsgd}
\eeq
\end{conj}

\bigskip

Conjecture \ref{conjsgd} and \ref{conjsgda} reduce to Theorems \ref{theosg2}, \ref{theosg3}, \ref{theosg4} and Corollaries \ref{corsg2a}, \ref{corsg3a}, \ref{corsg4a} respectively when $d$ is set to two, three and four.

By eq. (\ref{mcybound}), $z_{SG_d}$ has the upper bound $\ln b_{2d}$. It is of interest to see how close the value $z_{SG_d}$ is to this bound.  For this purpose, we define the ratio
\beq
r_{SG_d} = \frac{z_{SG_d}}{\ln b_{2d}} \ ,
\label{rupper}
\eeq
and list the first few values of $z_{SG_d}$ and $r_{SG_d}$ in Table \ref{ztable}. Our results agree with the observation made in Ref. \cite{sw} that $z_{SG_d}$ increases as the degree $k=2d$ increases. Compared with the values $z_{{\mathcal L}_d}$ for $d$-dimensional hypercubic lattice ${\mathcal L}_d$ which also has $k=2d$ in \cite{sw,Felker}, $z_{SG_d} < z_{{\mathcal L}_d}$ for all $d \ge 2$ indicates that $SG_d$ is less densely connected than ${\mathcal L}_d$. The ratio $r_{SG_d}$ increases with dimension and approaches one, but the convergence is not as fast as those for the regular lattices given in \cite{sw,sti}.

\begin{table}
\caption{\label{ztable} Numerical values of $z_{SG_d}$ and $r_{SG_d}$. The last digits given are rounded off.}
\begin{center}
\begin{tabular}{|c|c|c|c|c|}
\hline\hline 
$d$ & $D$ & $k$ & $z_{SG_d}$ & $r_{SG_d}$ \\ \hline\hline 
2  & 1.585 &  4 &  1.048594857  &  0.862051042 \\ 
3  & 2     &  6 &  1.569396409  &  0.928042816 \\ 
4  & 2.322 &  8 &  1.914853265  &  0.953722370 \\ 
5  & 2.585 & 10 &  2.172764568  &  0.966999152 \\ 
6  & 2.807 & 12 &  2.378271274  &  0.974945363 \\ 
7  & 3     & 14 &  2.548944395  &  0.980152960 \\ 
8  & 3.170 & 16 &  2.694814686  &  0.983785240 \\ 
9  & 3.322 & 18 &  2.822140640  &  0.986437211 \\ 
10 & 3.459 & 20 &  2.935085659  &  0.988442577 \\ 
\hline\hline 
\end{tabular}
\end{center}
\end{table}

\bigskip

Acknowledgments: The research of S.C.C. was partially supported by the NSC
grant NSC-95-2112-M-006-004 and NSC-95-2119-M-002-001. The research of L.C.C was partially supported by the NSC grant NSC-95-2115-M-030-002.

\vfill
\eject

\begin{thebibliography}{99}

\bibitem{kirchhoff}
G. Kirchhoff, \"Uber die Aufl\"osung der Gleichungen, auf welche man bei der Untersuchung der linearen Verteilung galvanischer Str\"ome gef\"uhrt wird, {\it Ann. Phys. Chem.} {\bf 72}: 497-508 (1847).

\bibitem{bbook}
N. L. Biggs, {\it Algebraic Graph Theory}, 2nd ed., Cambridge University Press, Cambridge, 1993.

\bibitem{welsh}
D. J. A. Welsh, {\it Complexity: Knots, Colourings, and Counting (London Math. Soc. Lecture Notes series 186)}, Cambridge University Press, Cambridge, 1993.

\bibitem{burton93} 
R. Burton and R. Pemantle, Local characteristics, entropy and limit theorems for spanning trees and domino tilings via transfer-impedances, {\it Ann. Probab.} {\bf 21}(3): 1329-1371 (1993).

\bibitem{lyons05}
R. Lyons, Asymptotic enumeration of spanning trees, {\it Combin. Probab. Comput.} {\bf 14}: 491-522 (2005).

\bibitem{temperley}
H. N. V. Temperley, The enumeration of graphs on large periodic lattices, in: D. J. A. Welsh and D. R. Woodall (Eds.), {\it Combinatorics: Proc. Combinatorial Mathematics}, The Institute of Mathematics and its Applications, Oxford, 1972, pp 356-357.

\bibitem{wu77}
F.-Y. Wu, Number of spanning tees on a lattice, {\it J. Phys. A: Math. Gen.} {\bf 10}(6): L113-L115 (1977).

\bibitem{fk}
C. M. Fortuin and P. W. Kasteleyn, On the random cluster model. I. Introduction and relation to other models, {\it Physica} {\bf 57}: 536-564 (1972).

\bibitem{wurev}
F.-Y. Wu, The Potts model, {\it Rev. Mod. Phys.} {\bf 54}(1): 235-268 (1982).

\bibitem{tzengwu}
W.-J. Tzeng and F.-Y. Wu, Spanning trees on hypercubic lattices and nonorientable surfaces, {\it Appl. Math. Lett.} {\bf 13}: 19-25 (2000).

\bibitem{sw}
R. Shrock and F.-Y. Wu, Spanning trees on graphs and lattices in $d$ dimensions, {\it J. Phys. A: Math. Gen.} {\bf 33}: 3881-3902 (2000).

\bibitem{std}
S.-C. Chang and R. Shrock, Some exact results for spanning trees on lattices, {\it J. Phys. A: Math. Gen.} {\bf 39}: 5653-5658 (2006).

\bibitem{sti}
S.-C. Chang and W. Wang, Spanning trees on lattices and integral identities, {\it J. Phys. A: Math. Gen.} {\bf 39}: 10263-10275 (2006).

\bibitem{mandelbrot}
B. B. Mandelbrot, {\it The Fractal Geometry of Nature}, Freeman, San Francisco, 1982.

\bibitem{Falconer}
K. J. Falconer, {\it Fractal Geometry: Mathematical Foundations and Applications}, 2nd ed., Wiley, Chichester, 2003.

\bibitem{Gefen80}
Y. Gefen, B. B. Mandelbrot and A. Aharony, Critical phenomena on fractal lattices, {\it Phys. Rev. Lett.} {\bf 45}: 855-858 (1980).

\bibitem{Gefen81}
Y. Gefen and A. Aharony, Solvable fractal family, and its possible relation to the backbone at percolation, {\it Phys. Rev. Lett.} {\bf 47}: 1771-1774 (1981).

\bibitem{Rammal}
R. Rammal and G. Toulouse, Spectrum of the Schr\"odinger equation on a self-similar structure, {\it Phys. Rev. Lett.} {\bf 49}: 1194-1197 (1982).

\bibitem{Alexander}
S. Alexander, Superconductivity of networks. A percolation approach to the effects of disorder, {\it Phys. Rev. B} {\bf 27}: 1541-1557 (1983).

\bibitem{Domany}
E. Domany, S. Alexander, D. Bensimon and L. P. Kadanoff, Solutions to the Schr\"odinger equation on some fractal lattices, {\it Phys. Rev. B} {\bf 28}: 3110-3123 (1983).

\bibitem{Gefen8384}
Y. Gefen, A. Aharony and B. B. Mandelbrot, Phase transitions on fractals: I. Quasi-linear lattices, {\it J. Phys. A: Math. Gen.} {\bf 16}: 1267-1278 (1983); Y. Gefen, A. Aharony, Y. Shapir and B. B. Mandelbrot, Phase transitions on fractals: II. Sierpinski gaskets, {\it ibid.} {\bf 17}: 435-444 (1984); Y. Gefen, A. Aharony and B. B. Mandelbrot, Phase transitions on fractals: III. Infinitely ramified lattices, {\it ibid.} {\bf 17}: 1277-1289 (1984).

\bibitem{Guyer}
R. A. Guyer, Diffusion on the Sierpinski gaskets: A random walker on a fractally structured object, {\it Phys. Rev. A} {\bf 29}: 2751-2755 (1984).

\bibitem{Kusuoka}
K. Hattori, T. Hattori and S. Kusuoka, Self-avoiding paths on the pre-Sierpinski gasket, {\it Probab. Theory Relat. Fields} {\bf 84}: 1-26 (1990); T. Hattori and S. Kusuoka, The exponent for the mean square displacement of self-avoiding random walk on the Sierpinski gasket, {\it ibid.} {\bf 93}: 273-284 (1992).

\bibitem{Dhar97}
D. Dhar and A. Dhar, Distribution of sizes of erased loops for lop-erased random walks, {\it Phys. Rev. E} {\bf 55}: R2093-R2096 (1997).

\bibitem{Daerden}
F. Daerden and C. Vanderzande, Sandpiles on a Sierpinski gasket, {\it Physica A} {\bf 256}: 533-546 (1998).

\bibitem{Dhar05}
D. Dhar, Branched polymers on the Given-Mandelbrot family of fractals, {\it Phys. Rev. E} {\bf 71}: 031801 (2005).

\bibitem{fh} 
F. Harary, {\it Graph Theory}, Addison-Wesley, New York, 1969.

\bibitem{mckay}
B. McKay, Spanning trees in regular graphs, {\it Europ. J. Combin.} {\bf 4} 149-160 (1983).

\bibitem{chungyau}
F. Chung and S.-T. Yau, Coverings, heat kernels and spanning trees, {\it J. Combin.} {\bf 6} 163-183 (1999).

\bibitem{Hilfer}
R. Hilfer and A. Blumen, Renormalisation on Sierpinski-type fractals, {\it J. Phys. A: Math. Gen.} {\bf 17}: L537-L545 (1984).

\bibitem{Felker}
J. L. Felker and R. Lyons, High-precision entropy values for spanning trees in lattices, {\it J. Phys. A: Math. Gen.} {\bf 36}: 8361-8365 (2003).

\end{thebibliography}
\end{document}